\documentstyle[epsfig]{aa}

\def\beq{\begin{equation}}
\def\enq{\end{equation}}
\def\bea{\begin{eqnarray}}
\def\ena{\end{eqnarray}}
\def\bec{\begin{center}}
\def\enc{\end{center}}

\def\ergcm2si{\hbox{ergs~cm$^{-2}$s$^{-1}$}}

\begin{document}

\title{The Analysis of Abell 1835 Using a Deprojection Technique}

\author{ S.M.~Jia\inst{1,2}, Y.~Chen\inst{1}, F.J.~Lu\inst{1}, L.~Chen\inst{1,2},
F.~Xiang\inst{1}}

\offprints{ Shumei~Jia (E-mail: jiasm@mail.ihep.ac.cn)}

\institute{Particle Astrophysics Center, Institute of High Energy Physics, 
Chinese Academy of Sciences, Beijing 100039, P.R. China    
\and
Department of Astronomy, Beijing Normal University, Beijing 100875, P.R. China }

\date{Received  / accepted }

\abstract{We present the results from a detailed deprojection analysis of Abell 
1835 as observed by {\it XMM-Newton}. If we fit the spectra with an isothermal 
plasma model, the deprojected temperature profile is flat in the outer 
region around 7.6 keV and decreases to $\sim$ 5.6 keV in the center, which may 
be connected with the gas cooling. In the central part, a two-component thermal 
plasma model can fit the spectrum significantly better. Moreover, the cool 
component (T $\sim$ 1.8 keV) has a much lower metal abundance than the 
hot component (T $\sim$ 8 keV), which may be due to the longer cooling time for
the cool gas with lower abundance. In addition, it was found that without a 
main isothermal component, the standard cooling flow model cannot fit the 
spectrum satisfactorily. From the isothermal model fitting results we also 
derived the electron density $n_e$, and fitted its radial distribution with a 
double-$\beta$ model.  The $n_e$ profile inferred with the double-$\beta$ model 
and the deprojected X-ray gas temperature profile were then combined to 
derive the total mass and the total projected mass of the cluster. The 
projected mass is lower than that derived from the weak lensing method. 
However, assuming that the cluster extends to a larger radius $\sim15'$ as found
by Clowe \& Schneider (2002), the two results are consistent within the error 
bars. Furthermore, we calculated the projected mass within the radius of $\sim$ 
153 kpc implied by the presence of a gravitational lensing arc, which is about 
half of the mass determined from the optical lensing. 
\\
}

\titlerunning{The Analysis of Abell 1835}
\authorrunning{Jia et al.}

\maketitle




\section{Introduction}
There are two yet not very well understood questions concerning clusters of 
galaxies. The first is: what is the real physical state and process of the X-ray 
gas in the central region of the cluster? The cooling flow rates in the centers 
of most galaxy clusters observed recently by {\it XMM-Newton} and {\it Chandra}
are much lower than those predicted by the standard cooling flow model (e.g. 
A1795, Ettori et al. 2002, Tamura et al. 2001; M87, Matsushita et al. 2002). 
An important implication of the low cooling flow rate is that there exist some 
unknown processes which can heat the gas and thus prevent it from cooling. 
From the analysis of the X-ray gas, some authors believe that the 
IntraCluster Medium (ICM) in some clusters is locally isothermal, such as in M87
(except for the regions associated with the radio structures, Molendi 2002), 
while some others suggested that the ICM can be better represented by a 
two-temperature model, in which the cooler component is associated with the 
Interstellar Medium (ISM) (Makishima et al. 2001). Apparently, a detailed 
analysis of the spectrum of the ICM in the cluster center is very helpful for 
exploring the actual physical state and the physical processes in the center. 
The second is: whether the cluster masses inferred from the two 
primary observational techniques (optical lensing and X-ray observation) are 
consistent? For some clusters, the mass determined from the 
X-ray method is lower than that from the strong gravitational lensing method by 
a factor $\sim$ 2 (e.g. A2218, Loeb \& Mao 1994; A1689, A2163, 
Miralda-Escud\'{e} \& Babul 1995). The reason for this discrepancy may be a 
prolate ellipsoidal mass distribution or a non-isothermal temperature structure 
existing in the cluster (Miralda-Escud\'{e} \& Babul 1995) or non-thermal 
pressure supporting the ICM against gravity (e.g. Loeb \& Mao 1994). 
Interestingly, when a multiphase analysis is adopted, the X-ray and strong 
lensing masses for the cooling flow clusters present an excellent agreement, 
but this discrepancy still exists in some other clusters which show no evidence 
for cooling flows (Allen 1998). However, since the standard multiphase model
does not give a satisfactory fit to the new {\it Chandra} and {\it XMM-Newton} 
data, it is necessary to explore whether this discrepancy do exist in the 
cooling flow clusters. Recently, Chen et al. (2003) found a discrepancy of a 
factor $\sim$ 2 in the cooling flow cluster PKS 0745-191 with {\it XMM-Newton} 
observations, while Schmidt et al. (2002) obtained a roughly consistent mass 
for the two methods with the {\it Chandra} data for Abell 1835. However, on 
larger spatial scales, previous studies have inferred a good agreement between 
the weak lensing mass and the X-ray mass (e.g. A2218, A2163, Squires et al 1996, 
1997).

Abell 1835 is an important object to study the above properties of the
galaxy clusters. It is luminous, with a medium redshift (z=0.2523) and a relaxed 
structure. {\it ROSAT} and {\it ASCA} observations show that it has a large 
cooling flow rate of $\sim1760^{+520}_{-590}$ M$_{\odot}$yr$^{-1}$ in its
center (Allen et al. 1996). However, recent studies based on  {\it XMM-Newton} 
(Turner et al. 2001) and {\it Chandra} (Weisskopf et al. 2000) show that the 
mass deposition rate is not that large. The RGS onboard {\it XMM-Newton} has 
limited the cooling flow rate to 315 M$_{\odot}$yr$^{-1}$ within a radius of 150 
kpc (Peterson et al. 2001), and {\it Chandra} derived that the cooling flow rate 
is $\sim500$ M$_{\odot}$yr$^{-1}$ within a radius of 250 kpc (Schmidt et al. 
2001). Furthermore, the {\it XMM-Newton} RGS did not detect any multiphase gas 
below a certain low temperature ($\sim$ 2.7 keV), inconsistent 
with the prediction of the standard cooling flow model (Peterson et al. 2001). 
The gas temperature of Abell 1835 was determined as $\sim$ 12 keV from the 
{\it Chandra} data (Schmidt et al. 2001), but was determined as $\sim7.6$ keV 
from the {\it XMM-Newton} data (Majerowicz et al. 2002). 
The discrepancy between the model prediction and observations as well 
as that between different observations urges a more careful examination with 
high quality data.   
 
We investigate here the temperature, density, cooling flow rate and mass of 
Abell 1835 with a deprojection technique based on the data observed by {\it 
XMM-Newton} EPIC. The deprojection technique can reveal the real spectra of the 
cluster gas in different spherical shells, and we can further determine the 
deprojected temperature and the mass distribution of the gas in the cluster 
(e.g. Chen et al. 2003). The {\it XMM-Newton} EPIC is the most sensitive X-ray 
telescope which also has high spatial and spectral resolutions, and therefore 
meets all the requirements for the detailed spectral analysis. 

The structure of this paper is as follows: Sect. 2 describes the 
observation, background correction and the spectral deprojection technique. 
Sect. 3 presents the deprojected spectral analysis with three different models: 
single-temperature model, two-temperature model and cooling flow model. 
In Sect. 4 we obtain the electron density profile, calculate the total mass, total 
projected mass and the projected mass within the optical lensing arc, then we 
discuss the discrepancy between the X-ray mass and the optical lensing mass. 
Our conclusion is given in Sect. 5.

Throughout this paper, the energy band is 0.5 $\sim$ 10 keV, and unless
otherwise noted we use a cosmology with 
$H_0$ = 50 km s$^{-1}${Mpc}$^{-1}$, $q_0$ = 0.5, $\Omega_m$ = 1.0, and 
$\Omega_{\Lambda}$ = 0. Therefore, 1$^{\prime}$ corresponds to 296.6 kpc at the 
distance of Abell 1835. 

\section{Observation and data preparation}
Abell 1835 was observed by {\it XMM-Newton} during the phase of performance 
verification (observation ID is 0098010101). Since MOS1 was operating in Large 
Window mode, we only use the data coming from MOS2 and pn cameras which were 
operating in the Full Frame mode and with the thin1 filter. The total exposure 
time is 60 ksec, of which only about 26 ksec is usable for spectroscopic 
analysis for both pn and MOS2. For the MOS2 data, we use the event with 
PATTERN $\le$ 12, and for the pn data PATTERN $\le$ 4. The calibration is 
performed in SAS 5.3.3. 

\subsection{Background correction}
Background subtraction was carried out using the same method as that of 
Majerowicz et al. (2002). The pn data have another source of contamination
called out-of-time (OOT) events counted during the read-out (see Str\"{u}der 
et al. 2001). This contamination has also been corrected for in our analysis.

\subsection{Spectral deprojection}
Abell 1835 appears to be a relaxed cluster of galaxies, therefore we assume 
that the temperature structure of this cluster is spherically symmetric, and
apply a deprojection technique as done by Nulsen \& B\"{o}hringer (1995). 
We divide the image of the cluster into 7 annular regions centered on the 
emission peak for the extraction of spectra and use the outmost ring
($8.33'-10.42'$) to determine the local Cosmic X-ray Background (CXB). However, 
in the sixth ring the signal-to-noise ratio is low, so we only consider the 
inner five regions ($r\le6'$). The minimum width of the rings was set to 
$0.75'$ which is wide enough to ignore the PSF (Point Spread Function), whose 
FWHM (Full Width at Half Maximum) is 5$\arcsec$ for MOS2 and 6$\arcsec$ for pn. 
For each annular region, an Ancillary Response File (ARF) is generated using 
SAS, then through the ARF, the vignetting (Arnaud et al. 2001) correction is 
administered. 

Deprojected spectra are calculated by subtracting the contribution from the
outer regions for all spectral components. Within each annular region, the 
spectrum per unit volume is assumed to be the same. The deprojected spectrum 
of the $i$th shell is then calculated by subtracting the contributions from 
$i$+1th to the outmost shell from the annular spectrum of the corresponding 
radius (e.g. Matsushita et al. 2002).

\section{Spectral analysis}
\subsection{Single temperature model}
We analysed the deprojected spectra of both MOS2 and pn data using XSPEC 
version 11.2.0 (Arnaud 1996), and we selected the following model:
\begin{equation}
Model_1=Wabs(n_H)\times Mekal(T,z,A,norm),
\end{equation}
here Wabs is a photoelectric absorption model  (Morrisson \& McCammon 1983) 
and Mekal is a single temperature plasma emission model (Mewe et al.1985, 1986; 
Kaastra 1992; Liedahl et al. 1995). We fixed the redshift $z$ to 0.2523 and $n_H$ 
to the Galactic absorption 2.24$\times$10$^{20}$ cm$^{-2}$ (Dickey \& Lockman 
1990). The results are listed in Table 1 and the spectrum of the central region 
(r $<0.75'$) for MOS2 data fitted by this single temperature model is shown as 
Fig.3 (a).

\begin{table*}
\caption{The best-fit free parameters of Abell 1835: the temperature T of 
MOS2, pn and the combination of them; the abundance A and the normalized 
constant $norm$ for the  combination of MOS2 and pn. $norm = 10^{-14}/(4\pi D^2)
\int n_e n_H dV$, where $D$ is the distance to the source (cm) and $n_e$ is the 
electron density (cm$^{-3}$). $L_X$ is the bolometric luminosity and P is the null 
hypothesis probability of the spectra fitted in XSPEC. The errors represent a 
confidence level of 90\%.}
\begin{center}
{\footnotesize
\begin{tabular}{c@{\hspace{0.15cm}}c@{\hspace{0.15cm}}c@{\hspace{0.15cm}}
c@{\hspace{0.15cm}}c@{\hspace{0.15cm}}c@{\hspace{0.15cm}}c@{\hspace{0.15cm}}c}
\hline
\multicolumn{2}{c}{Annulus (')} & \multicolumn{3}{c}{T (keV) 
($\chi^{2}_{\mathrm{red}}$/$dof$,P)} & A (solar)& norm($10^{-3}$cm$^{-5}$)
& $L_X$($10^{45}$erg s$^{-1}$)
\\
$r_{1}$ & $r_{2}$ & {\sc mos}2 & pn & {\sc mos}2+pn & {\sc mos}2+pn &
{\sc mos}2+pn & (0.001-60 keV)
\\
\hline
0.0 & 0.75 & $6.24\pm0.28$  & $5.46\pm0.10$  & 
$5.60\pm0.10$  & $0.36\pm0.03$ & $9.02\pm0.08$ & 3.62
\\
  &  & (1.31/154, 0.006) & (1.56/445, 3$\times10^{-13}$) &
  (1.53/602, 4$\times10^{-16}$) & & &
\\
0.75 & 1.5 & $8.37^{+0.86}_{-0.84}$  & $7.39^{+0.53}_{-0.35}$ 
 & $7.71^{+0.42}_{-0.40}$  & $0.22\pm0.07$ & $4.26\pm0.07$ & 1.90
\\
  &  & (1.03/101, 0.395) & (1.04/278, 0.293) & (1.21/382, 0.003) & & &
\\
1.5 & 2.25 & $7.82^{+1.60}_{-1.28}$  & 
$7.16^{+0.79}_{-0.59}$  & $7.27^{+0.77}_{-0.51}$ &
$0.23\pm0.12$ & $2.45\pm0.07$ & 1.07
\\
  & & (1.10/59, 0.278) & (1.01/168, 0.430) & (1.04/230, 0.338) & & &
\\
2.25 & 3.33 & $7.93^{+1.89}_{-1.30}$   & $7.10^{+1.20}_{-0.95}$ 
& $7.33^{+1.05}_{-0.76}$  & $0.35\pm0.18$ & $1.55\pm0.07$ & 0.70
\\
  &  & (1.26/45, 0.10) & (0.84/108, 0.885) & (1.18/156, 0.06) & & &
\\
3.33 & 6.0 & $7.32^{+7.31}_{-2.84}$  & $6.61^{+1.76}_{-1.31}$ 
 & $6.63^{+1.52}_{-1.20}$  & $0.36^{+0.39}_{-0.32}$ & $1.42\pm0.12$ & 0.61
\\
  &  & (0.84/27, 0.80) & (1.26/116, 0.03) & (1.37/146, 0.002) & & &
\\
\hline
\end{tabular}
}
\end{center}
\label{tablefit}
\end{table*}

To derive the temperature profile of this cluster, we fitted the
combined spectra of MOS2 and pn with a single temperature model; the results 
are also listed in Table 1. From the deprojected temperature profile (the 
diamonds in Fig.1), we can see that the temperature is nearly constant in the 
outer regions but decreases towards the center, which may be connected with the 
gas cooling. Then we fit the temperature profile with the formula:
\begin{equation}
T(r)=ae^{br}+c.
\end{equation}
The best fit parameters are: $a$ = -101.9 keV, $b$ = -0.11 arcsec$^{-1}$, 
$c$ = 7.55 keV.

Our result does not differ much from that of Majerowicz et al. (2002) 
based on the same data, although what Majerowicz et al. derived is the projected
temperature. However, the deprojected temperature from the {\it Chandra} data
(Schmidt et al. 2001), $\sim$ 12 keV, is much higher than our result. This may 
be due to the difficulty in identifying the background flares in the {\it 
Chandra} data (Markevitch 2002). 

\begin{figure}[ht]
\centerline{\psfig{file=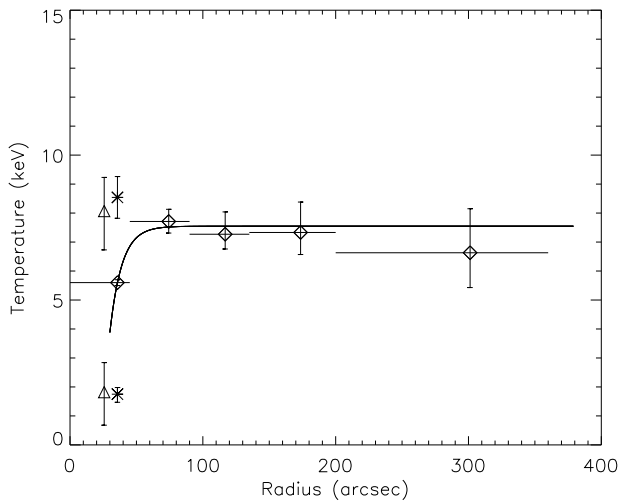,width=9cm}}
\caption{Temperature profile of Abell 1835 for the combination of 
{\sc mos}2 and pn with a confidence level of 90\%. (diamonds: temperature from 
the 1T model fitting of the combined spectra; stars: temperature from the 2T 
model fitting of the pn central spectrum; triangles: temperature from the 2T 
model fitting of the MOS2 central spectrum. We have offset the 
triangles $10''$ to the left so as to illustrate the two fits clearly.) 
The solid line is the best-fitted profile of the 1T model.}
\end{figure}

\subsection{Double temperature model}
From the null hypothesis probability P (the adopted level of a fit is
$P\ge0.1$, and when $0.01\le P<0.1$, the fit is marginally acceptable) in 
Table 1 we find that the spectral fitting with the single-temperature model is
acceptable for the outer regions, but unacceptable for the central region.
Since the calibrations of pn and MOS2 are not quite consistent with each other,  
we fit the central spectrum of pn and MOS2 with the two-temperature model 
respectively:
\begin{eqnarray}
Model_2&=&Wabs(n_H)\times(Mekal(T_1,z,A_1,norm_1)+ \nonumber \\  
  && Mekal(T_2,z,A_2,norm_2)),
\end{eqnarray}
which means that there are two components in this region with different
temperatures, and the free parameters have the same meaning as in Sect. 3.1. 
Assuming that the two components have the same abundances, we fix the 
abundances of the cool component to those of the hot gas (see Fig.3 (b) for
MOS2), and obtain the best-fit parameters listed in Table 2. Note that the 
cool component only has a small volume filling fraction of less than a 
few percent for both pn and MOS2 (see $f_{vol}$ in Table 2). Although the best
parameters derived from pn and MOS2 do not agree with each other in detail, 
the values of $\chi^{2}$ for both of them improved significantly and are marginally
acceptable.
 
We further leave the abundances free and obtain a even better fit (see Fig.3 (c)).
The resultant model parameters of pn and MOS2 are now consistent with each
other, as shown in Table 3, where $F$ is derived from a simple $F$-test 
(Bevington 1969) and indicates the degree of confidence of the model in 
which both abundances are left free parameters compared to
the model setting the two abundances the same. From $F$ we 
know that the substitution of this model for pn data is significant with $>$ 
99.9\% confidence, while for MOS2 it is about 80\%. The existence of the two 
abundance components implies that the distribution of the abundance for both 
temperature components is inhomogeneous in the central region of the cluster. 
As shown in Table 3, in the steady cooling flow scenario, there exist two 
temperature components, and the cool one with a temperature less than 2 keV 
has a lower abundance. This may be due to the metal-rich gas cooling 
together with the metal-poor gas by bremsstrahlung emission until $\sim$2 keV, 
below which line cooling becomes important, and the metal-rich gas would cool 
at a much faster rate than the metal-poor gas (Fabian et al. 2001). So in the 
steady cooling flow scenario the metal-rich component cannot be seen at 
temperatures below about 2 keV and thus the cool component presents a lower 
abundance.

\begin{table*}
\caption{The best-fit parameters of Abell 1835 using a two-temperature 
model for the central region($0.0'-0.75'$) and assuming that the two components 
have the same abundance. The error bars are at the 90\% confidence level. 
$f_{vol}$ is the volume fraction of the cool component.}
\begin{center}
\begin{tabular}{ccccccccc}
\hline
Detector & \multicolumn{2}{c}{T (keV)} & A & 
\multicolumn{2}{c}{norm($10^{-3}$cm$^{-5}$ )} & $f_{vol}$ & 
$\chi^{2}_{\mathrm{red}}$/$dof$ & P
\\
   & $T_1$ & $T_2$ & (solar) & $norm_1$ & $norm_2$ &  &  &
\\
\hline
pn & $8.17^{+1.01}_{-0.77}$  & $2.02\pm0.28$  & 
$0.35\pm0.05$  & $6.67^{+0.60}_{-0.71}$ & $2.59^{+0.73}_{-0.64}$ & 0.02 &
1.21/443 & 0.02
\\
MOS2 & $6.60\pm0.33$ & $0.71\pm0.18$ & 
$0.44\pm0.07$  & $8.67^{+0.10}_{-0.17}$ & $0.16\pm0.08$ & 0.0002
& 1.24/152 &  0.02
\\
\\
combined & $8.05^{+0.82}_{-0.70}$  & $2.12^{+0.33}_{-0.34}$  & 
$0.37\pm0.04$ & $6.84^{+0.63}_{-0.66}$ & $2.35^{+0.67}_{-0.65}$ & 0.02 &
1.29/600 & 2$\times 10^{-6}$ 
\\
\hline
\end{tabular}
\end{center}
\label{tablefit}
\end{table*}

\begin{table*}
\caption{The best-fit parameters of Abell 1835 using a two-temperature 
model for the central region ($0.0'-0.75'$), but leaving the abundances of the 
two components free. The error bars are at the 90\% confidence level. $F$ is 
the confidence level of using this 2T2A model instead of the 2T1A model.}
\begin{center}
\begin{tabular}{cccccccccc}
\hline
Detector & \multicolumn{2}{c}{T (keV)} &
\multicolumn{2}{c}{A (solar)} & \multicolumn{2}{c}{norm($10^{-3}$cm$^{-5}$)} & 
$\chi^{2}_{\mathrm{red}}$/$dof$ & P & $F$
\\
 & $T_1$ & $T_2$ & $A_1$ & $A_2$ & $norm_1$ & $norm_2$ & & &
\\
\hline
pn & $8.54\pm0.72$  & $1.75^{+0.23}_{-0.28}$  & 
$0.50\pm0.08$  
& $0.13\pm0.06$ & $6.39^{+0.55}_{-0.50}$ & $3.34\pm0.50$ &
1.13/442 &  0.03 & $>$ 0.999
\\
MOS2 & $8.07^{+1.16}_{-1.34}$  & $1.82^{+1.02}_{-1.14}$  &
$0.52^{+0.14}_{-0.11}$ & $0.10^{+0.21}_{-0.10}$ & $7.36^{+1.30}_{-1.17}$ & 
$2.08^{+0.98}_{-1.76}$ & 1.23/151 & 0.03 & 0.80
\\
\\
combined & $8.32^{+0.60}_{-0.65}$  & $1.75^{+0.23}_{-0.31}$  &
$0.49\pm0.07$ & $0.12\pm0.05$ & $6.69^{+0.52}_{-0.44}$ & 
$3.0^{+0.44}_{-0.48}$ & 1.22/599 & 0.0001 & $>$ 0.999  
\\
\hline
\end{tabular}
\end{center}
\label{tablefit}
\end{table*}

Although the combined spectra of pn and MOS2 are not suitable for the 
two-temperature model fitting because of the slight inconsistence between their
calibrations, we still perform the combined fits to calculate the mass in the 2T 
model method easily (see Sect. 4.4). The resulting parameters as listed in 
Table 2 and 3 lie between the two sets of parameters derived separately from 
the pn and MOS2 data. 

We also fit the outer spectra with the two-temperature model, but the 
fits do not improve significantly.

From the analyses above, we know that a two-temperature model may represent the 
spectrum of the central region much better, which is possibly due to the 
existence of cooling gas in the cluster center or the possible 
presence of ISM associated with the cD galaxy (Makishima et al. 2001). However, 
for the outer parts, the fit of the single-temperature model is good enough 
to be acceptable. So we inferred that in the cluster center there exist two 
gas components with different temperatures, while in the outer regions the gas 
is isothermal (Kaastra et al. 2003). 

Fig.2 is the deprojected abundance profile. It can be seen that the abundance 
is higher in the cluster center and tends to constant in the outer region. This 
indicates that the excess metal in the cluster center is produced in the cD 
galaxy and ejected into the ICM (Makishima et al. 2001). 
 
\begin{figure}[ht]
\centerline{\psfig{file=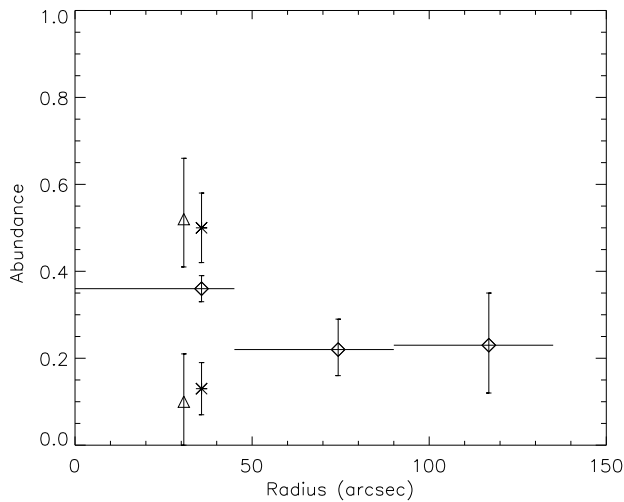,width=9cm}}
\caption{Abundance profile of Abell 1835 in units of solar metallicity with a 
confidence level of 90\%. The symbols have the same meaning as Fig.1. }
\end{figure}

\subsection{Asorbing column density}
To verify if the observed $n_H$ is consistent with the 
Galactic column density, we fit the spectra with the parameter $n_H$ left free.
First, we perform joint fits to 5 pn and MOS2 spectra respectively with a single
temperature model, assuming that the 5 spectra have the same $n_H$, and the
results are shown in the upper part of Table 4. It can be seen that the two 
best-fit $n_H$ are much smaller than the Galactic column density $\sim$
$2.24\times10^{20}$ cm$^{-2}$. Since the central spectrum can be fitted well by 
a two-temperature model, we consider a joint fit with a two-temperature
model for the central spectrum and a single temperature model for the outer 4
spectra, and the hydrogen column density $n_H$ of each spectrum is assumed to be
the same. The results are listed in the lower part of Table 4, where $F$ is the 
confidence level corresponding to using the 2T model to fit the central 
spectrum instead of using the 1T model to fit all five spectra. We find that 
the fits become better, and from $F$ we know that the substitution by a 2T 
model is necessary. It is also shown in Table 4 that the $n_H$ determined from 
the pn spectra is still smaller and about half the Galactic column density, 
while the $n_H$ obtained from the MOS2 spectra is consistent with the Galactic 
column density. We should note that for many clusters the best-fit column 
density derived from the MOS is always in good agreement with the Galactic 
value, while that estimated from pn data is systematically smaller (Molendi \& 
Pizzolato 2001). This may be due to the MOS detector having a more reliable 
calibration than the pn detector.


We also note here that the joint fit with a two-temperature model for the
central spectrum can yield a much more believable $n_H$, and this is also 
evidence for the existence of two gas components with different temperatures in 
the cluster center.

\begin{table}
\caption{The absorbing column density determined by pn and MOS2 spectra assuming
that all the spectra have the same $n_H$. 1T model: the joint fit with a single 
temperature model. 2T+1T model: the joint fit with a two-temperature model for 
the central spectrum and a single temperature model for the outer 4 spectra, 
where $F$ is the confidence level of the substitution by the 2T+1T model.}
\begin{center}
\begin{tabular}{c@{\hspace{0.15cm}}c@{\hspace{0.15cm}}c@{\hspace{0.15cm}}
c@{\hspace{0.15cm}}c@{\hspace{0.15cm}}c}
\hline
 Model & Detector & $n_H$ & $\chi^{2}_{\mathrm{red}}$/$dof$ & P & $F$
 \\
 & & ($10^{20}$cm$^{-2}$) &  &  & 
\\
\hline
 1T & pn & $0.01^{+0.30}_{-0.01}$ & 1.15/1114 & 0.0004 &
\\
 & MOS2 & $1.05^{+0.65}_{-0.62}$ &  1.15/385 & 0.02 & 
\\
\\
 2T+1T & pn & $0.99^{+0.45}_{-0.43}$ & 1.06/1111 & 0.08 & $>$0.999
\\
 & MOS2 & $1.97^{+1.12}_{-0.98}$ &  1.13/382 & 0.035 &  $>$0.90
\\
\hline
\end{tabular}
\end{center}
\label{tablefit}
\end{table}

Through the analysis above, we can conclude that the observed $n_H$ of {\it
XMM-Newton} is consistent with the Galactic column density, and so 
the analysis in Sect. 3.1 and Sect. 3.2 is reliable. In addition, the MOS data 
are probably much better for determining $n_H$ than the pn data.

\subsection{Cooling flow model}
X-ray observations of clusters of galaxies show that in the central regions of
some clusters the cooling time of the ICM is significantly less than the Hubble 
time (Edge et al 1992; White, Jones \& Forman 1997; Peres et al. 1998). ROSAT
and ASCA observations show that Abell 1835 is the cluster that contains
the largest cooling flow rate (Allen et al. 1996). Here we use the MOS2 data, 
which are assumed to be better for the determination of $n_H$ and thus the 
cooling flow rate (Molendi \& Pizzolato 2001), to calculate the cooling flow 
rate of Abell 1835. There are two different methods for calculating the cooling 
flow rate: spectral method and spatial method. 

Since we have already deprojected the spectrum, i.e., the contribution of the
ambient gas has been excluded, the spectrum should be fitted well with the
standard cooling flow model:
\begin{eqnarray}
Model_3 = Wabs(n_H)\times(Zwabs(\Delta n_H)\times Mkcflow(\dot{M})),
\end{eqnarray}
where Wabs was described in Sect. 3.1, Zwabs is an intrinsic photoelectric 
absorption model (Morrison \& McCammon 1983), and Mkcflow is a cooling flow 
model (Fabian, 1988); $\Delta n_H$ is the intrinsic absorption and $\dot{M}$ 
the rate of gas cooling out of the flow. Since the cooling flow radius of Abell
1835 is $\sim$ 230 kpc (Allen 2000) at which the cooling time first exceeds the 
age of the universe, we only fit the central spectrum ($\sim$ 225 kpc) with this
model (Table 5). It can be seen that to fit 
the spectrum a cut-off temperature, low$T$, is needed, which is found to be 
2.3 keV. If we fix low$T$ = 0.01 keV, the $\chi^{2}$ value will be much larger
(see Table 5) and from P we know that this fit is unacceptable. Therefore, 
the standard cooling flow model cannot fit the spectrum satisfactorily.

\begin{table*}
\caption{The best-fit parameters for the central region (r $<0.75'$) of MOS2 
data by the standard cooling flow model. 
The errors are at the 90\% confidence level. $\dot{M}$ is the mass deposition rate, 
and $\Delta n_H$ is the intrinsic absorption.}
\begin{center}
\begin{tabular}{ccccccc}
\hline
 low$T_{cf}$ & high$T_{cf}$ & A
  & $\dot{M}$ & $\Delta n_H$ & $\chi^{2}_{\mathrm{red}}$/$dof$ & P
\\
 (keV) & (keV)& (solar) & (M$_{\odot}$) & ($10^{22}$cm$^{-2}$) & &
\\
\hline
 $2.27^{+0.84}_{-0.58}$ & $13.75^{+2.76}_{-2.62}$ & $0.46\pm0.08$ &
$2527.4^{+967.9}_{-488.7}$ & $0.0^{+0.003}_{-0.0}$ & 1.25/152 & 0.02
\\
 0.01 (fix) & $18.21^{+1.88}_{-1.31}$ & $0.35\pm0.08$ &
 $1766.6^{+118.7}_{-141.0}$ & $0.06^{+0.02}_{-0.01}$ & 1.37/153 & 0.001
\\
\hline
\end{tabular}
\end{center}
\label{tablefit}
\end{table*}

When adding an isothermal Mekal component, the fit becomes much better, as
shown in Table 6. $F$ shows that the replacement ((Mekal+Mkcflow) instead of
Mkcflow) is significant, i.e., the cooling flow model with a Mekal component is
more reliable, which implies that there should exist an isothermal component 
in the cluster. Then, the cooling flow rate 
$\dot{M}$ is about 656.2$^{+403.4}_{-360.2}$ M$_{\odot}$yr$^{-1}$ within 225 
kpc ($0.75'$). The spectrum of the central region ($0.0'-0.75'$) fitted by this 
model for MOS2 data is shown in Fig.3 (d).

\begin{table*}
\caption{The same as Table 5 but for the cooling flow model with a Mekal
component. $F$ is the confidence level when using (Mekal+Mkcflow) instead of
the standard cooling flow model.}
\begin{center}
\begin{tabular}{cccccccccc}
\hline
 $T_{mekal}$ & low$T_{cf}$ & high$T_{cf}$ & A 
  & norm & $\dot{M}$ & $\Delta n_H$ & $\chi^{2}_{\mathrm{red}}$/$dof$ & P & $F$
\\
 (keV) & (keV) & (keV)& (solar) & ($10^{-3}$cm$^{-5}$) & 
(M$_{\odot}$) & ($10^{22}$cm$^{-2}$) &  &  &
\\
\hline
 $7.30^{+0.91}_{-0.60}$ & 0.01 (fix) & =$T_{mekal}$ & $0.44\pm0.08$ &
$7.20^{+0.92}_{-1.13}$ & $656.2^{+403.4}_{-360.2}$  & 
$0.06\pm0.06$ &  1.22/152 & 0.03 & $>$0.999
\\
$6.38^{+0.36}_{-0.38}$ & 0.01 (fix) & =$T_{mekal}$ & $0.43\pm0.07$ &
$8.41^{+0.63}_{-0.85}$ & $262.5^{+169.5}_{-262.5}$& 0.25 (fix) & 
1.31/153 & 0.006&
\\
\hline
\end{tabular}
\end{center}
\label{tablefit}
\end{table*}

It can also be seen from Table 6 that the intrinsic absorption $\Delta{n_H}$ of 
Abell 1835 is very small and consistent with zero within the error bars, which 
is in agreement with the previous results, such as those of RGS (Peterson et al. 
2001) and Molendi \& Pizzolato (2001) in which the projected spectra are used. 
If we adopt the best-fit result from the analysis in Schmidt et al. (2001) of 
the Chandra dataset, and fix $\Delta{n_H}=0.25\times10^{22}$ cm$^{-2}$, we 
measure $\dot{M}=262.5^{+169.5}_{-262.5}$ M$_{\odot}$yr$^{-1}$, which is 
unacceptable at the adopted level of P=0.01. So the intrinsic absorption of 
Abell 1835 is very small and close to zero. 


Another method for calculating the cooling flow rate is the spatial method.
According to the energy conservation, this method can be expressed as (White,
Jones \& Forman 1997):
\begin{eqnarray}
L_x(i)&=&\dot{M}(i)[h(i)+f(i)\bigtriangleup\phi(i)]+ \nonumber \\  
  && \sum_{i'=1}^{i'=i-1}\dot{M}(i')[\bigtriangleup h(i)+\bigtriangleup\phi(i)],
\end{eqnarray}
where $L_x$ is the bolometric luminosity, shown in Table 1; $\dot{M}(i)$ is the mass
deposition in shell $i$; $\sum_{i'=1}^{i'=i-1}\dot{M}(i')$ is the mass of gas
that needs to pass through shell $i$ to give rise to the radiation and mass
deposition in interior shells; $\bigtriangleup\phi(i)$ is the change in the
gravitational potential; $h(i)=\frac{5}{2}kT(i)/\mu m_p$ is the
temperature in units of energy per particle mass of the hot gas; $f(i)$ is a
fraction of the overall change in the cluster potential, and in this analysis
$f(i)=1$ is used. The first two terms in this equation represent the mass that 
is left in shell $i$, and the second two terms represent the mass that flows 
through to the next interior shell.
From this method, we can estimate the cooling flow rate within the 
central region ($r<225$ kpc) $\sim$ about 1600 M$_{\odot}$yr$^{-1}$, consistent 
with the previous result of {\it Chandra} from the same method (Schmidt et al. 
2001).

The result derived from the spatial method gives the total loss rate of 
the energy, but the cooling flow rate obtained from the spectral method only 
contains the mass deposition rate during the cooling flow. The discrepancy 
between these two methods implies that there should exist some other energy 
sources, for example heating by AGN (see e.g. B\"{o}hringer et al. 2002) or 
thermal conduction (e.g. Voigt et al. 2002) to support the X-ray emission in 
the center of the cluster.

With any acceptable model ( the two-temperature model or the cooling flow 
model with a mekal component) we have investigated above, we can conclude that 
in the cluster center there should always exist a locally isothermal component 
and there should also exist a minor component which may be a single phased cool
component or a multiphased one. However, with the current data we can not
discriminate between these two models.

\begin{figure*}
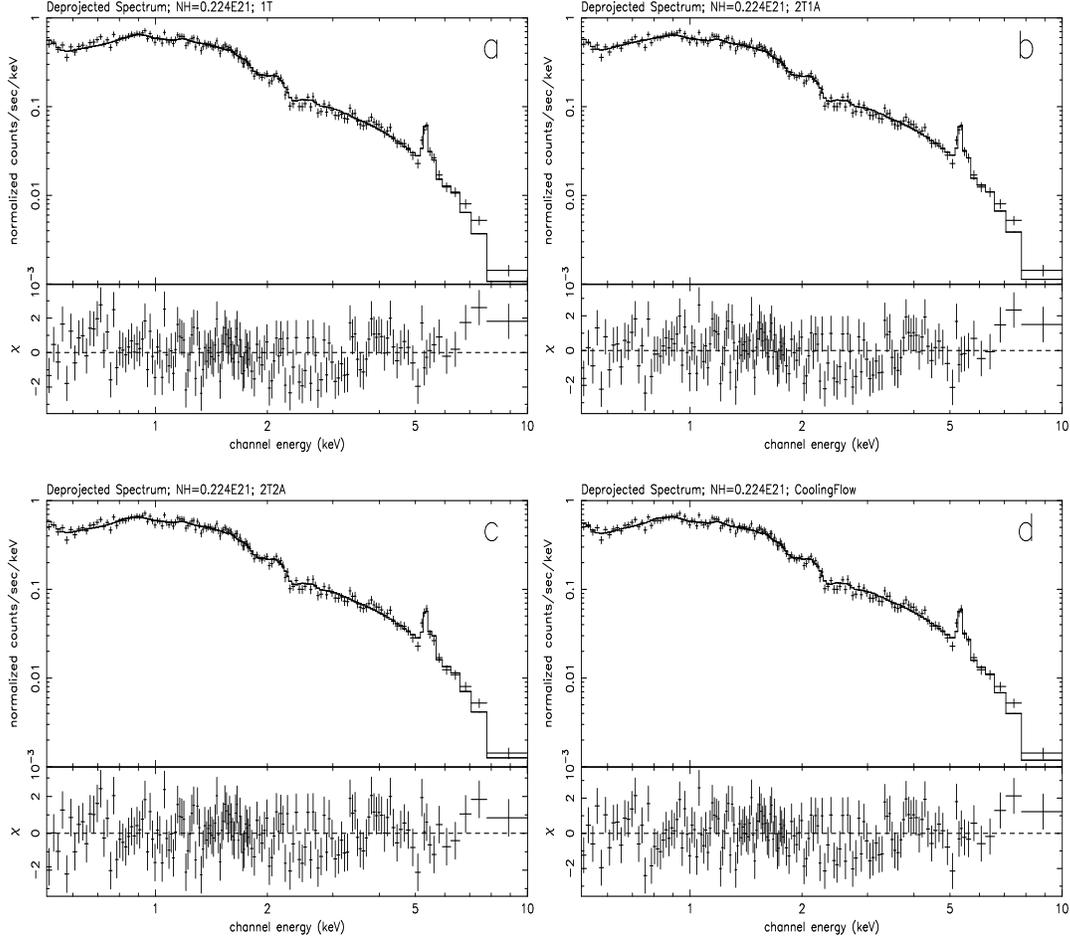

\centerline{\hbox{\psfig{figure=0006fig3.ps,width=6cm,height=7cm,angle=-90}
\psfig{figure=0006fig4.ps,width=6cm,height=7cm,angle=-90}}} 
\hbox{}
\centerline{\hbox{\psfig{figure=0006fig5.ps,width=6cm,height=7cm,angle=-90}
\psfig{figure=0006fig6.ps,width=6cm,height=7cm,angle=-90}}} 
\hbox{}
\caption {The spectrum of the central region (r $<0.75'$) for MOS2 data of Abell 
1835. a) fitted by the 1T model; b) fitted by the 2T1A model; c) 
fitted by the 2T2A model; d) fitted by the cooling flow model 
(wabs(mekal+zwabs(mkcflow))).}
\end{figure*}

\section{Mass analysis}
\subsection{Electron density}
Here, we have divided the cluster into 17 annular regions centered on the 
emission peak. By the deprojection technique we can calculate the photon counts 
in each corresponding shell. Then, using the abundance and the temperature 
profile, the normalized constant of each region, $norm$, can be estimated. 
Furthermore, since:
\begin{equation}
norm = 10^{-14}/(4\pi D^2)\int n_e n_H dV,
\end{equation}
we can derive the deprojected electron density $n_e$, shown as the triangles 
in Fig.4. 

We fit the electron density with both single-$\beta$ and double-$\beta$
model. First, we fit it with a single-$\beta$ model (e.g. Cavaliere \& 
Fusco-Femiano 1976) in which the electron density profile $n_e(r)$ is defined 
as:
\begin{equation}
n_e(r)=n_{0} {\left[1+{\left(\frac{r}{r_{c}}\right)}^2\right]}^{-{\frac{3}{2}}\beta},
\end{equation}
where $n_0$ is the central density, $r_c$ is the core radius, and $\beta$ is
the slope. The best-fit parameters are: $n_0$ = 0.06 cm$^{-3}$, 
$r_c$ = 0.21 arcmin, $\beta$ = 0.56, $\chi^{2}$=60.33, DOF=14 and the 
best-fit single-$\beta$ model profile is plotted as a dotted line in 
Fig.4. We can see that it does not fit well, especially in the outer regions. 
For this reason we fit it with a `double-$\beta$' model (see Chen et al. 2003):
\begin{equation}
n_e(r)=n_{01} {\left[1+{\left(\frac{r}{r_{c1}}\right)}^2\right]}^{-{\frac{3}{2}}\beta_1}+n_{02}
{\left[1+{\left(\frac{r}{r_{c2}}\right)}^2\right]}^{-{\frac{3}{2}}\beta_2}.
\end{equation}
Note that this model fits the electron density directly and is different from 
the common double-$\beta$ model that was used to fit the surface brightness. 
The best fit parameters are: $n_{01}$ = 0.003 cm$^{-3}$, $r_{c1}$ = 1.84 arcmin, 
$\beta_1$ = 1.24, $n_{02}$ = 0.08 cm$^{-3}$, $r_{c2}$ = 0.155 arcmin, 
$\beta_2$ = 0.60, $\chi^{2}$=12.25, DOF=11 and the best-fit line is shown as 
the solid line in Fig.4. It can be seen from the $\chi^{2}$ value that the 
electron density profile is fitted better by a double-$\beta$ model than by a 
single-$\beta$ model, so in the following calculations we use the electron 
density profile inferred from the double-$\beta$ model.

\begin{figure}[ht]
\centerline{\psfig{file=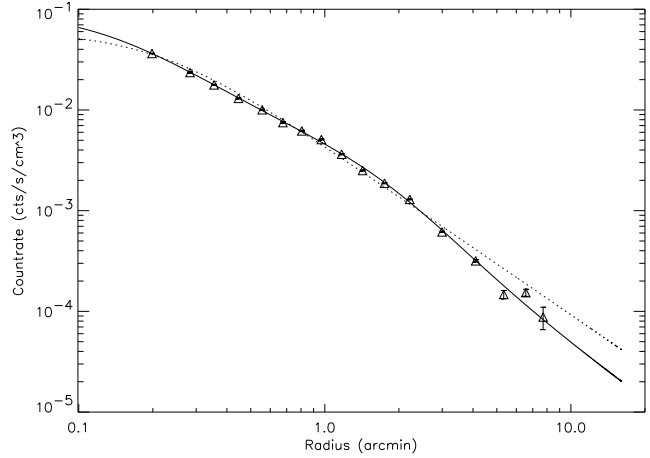,width=9cm}}
\caption{Electron density profile of Abell 1835. The error bars are at a 68\% 
confidence level. The solid line is the best-fit profile for a 
double-$\beta$ model, and the dotted line for a single-$\beta$ model.}
\end{figure}

\subsection{Total mass}
Assuming spherical symmetry and hydrostatic equilibrium, the total mass profile 
can be determined when the radial profiles of the gas density and temperature 
are known. We calculate the gravitational mass of Abell 1835 with the 
hydrostatic equation (Fabricant et al. 1980):
\begin{equation}
M_{tot}(<r)=-\frac{k_B T r^2}{G\mu
m_p}[\frac{d(\ln{n_e})}{dr}+\frac{d(\ln{T})}{dr}],
\end{equation}
here $k_B$ is the Boltzmann constant, $G$ is the gravitational constant, $\mu$ 
is the mean molecular weight of the gas in units of $m_p$, the proton mass. 
For a fully ionized gas with a standard cosmic abundance, 
a suitable value is $\mu$ = 0.6.

Using the best-fit profiles of the electron density $n_e(r)$ and the 
deprojected temperature $T(r)$ we can obtain the total mass profile shown in 
Fig.5. It was found that the total mass $M_{tot}$ within the radius of $6'$ is 
$1.05\pm0.13\times10^{15}$ M$_{\odot}$, which is consistent with that found by 
Majerowicz et al. (2002) within the errors. 

\begin{figure}[ht]
\centerline{\psfig{file=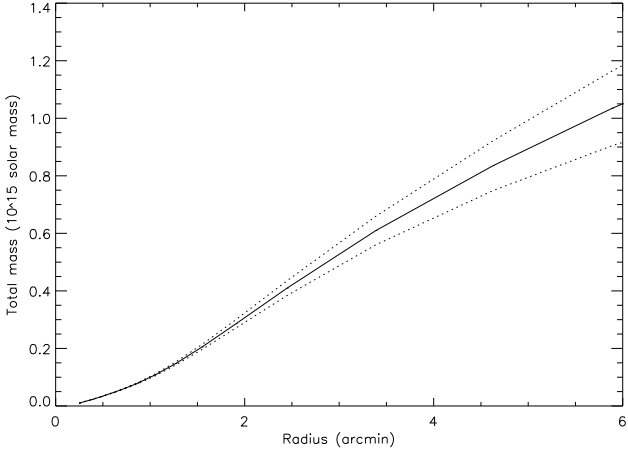,width=9cm}}
\caption{The total mass profile of Abell 1835. The error bars (dotted lines)
are at a 68\% confidence level.}
\end{figure}

\subsection{Total projected mass}
We calculate the projected mass profile of Abell 1835 for another universe model:
$H_0$ = 100 km s$^{-1}${Mpc}$^{-1}$, $\Omega_m$ = 0.3, and $\Omega_{\Lambda}$ 
= 0.7 and compare it with that of Clowe \& Schneider (2002), who also 
calculated the total mass in the same universe model using the weak lensing 
method based on the data from the ESO/MPG Wide Field Imager (see Fig.6). The 
upper panel shows our projected mass within $6'$, and it can be seen 
that the two results are consistent within our error bars in the radial 
range between 1' and 4', where lensing mass estimates are available. By 
increasing the radius, our mass measurement becomes lower than that of Clowe 
\& Schneider. Since Clowe \& Schneider have derived the lensing mass profile up 
to $15'$, we also extrapolate our best-fit temperature and electron density 
profiles to that large radius, then calculate the projected mass, shown in the 
lower panel. The two results are consistent within our error bars. Moreover, 
if we assume that the mass in the outer regions is distributed as derived 
from Clowe \& Schneider, we estimate a contribution projected from $r>6'$ to 
the inner part of about $0.16\times10^{15}$ M$_{\odot}$. By considering this
contribution, the two results in the upper panel will also be consistent. 
Therefore we conclude that the X-ray mass shows an excellent agreement with 
the weak lensing mass at large radii.

\begin{figure}[ht]
\centerline{\psfig{file=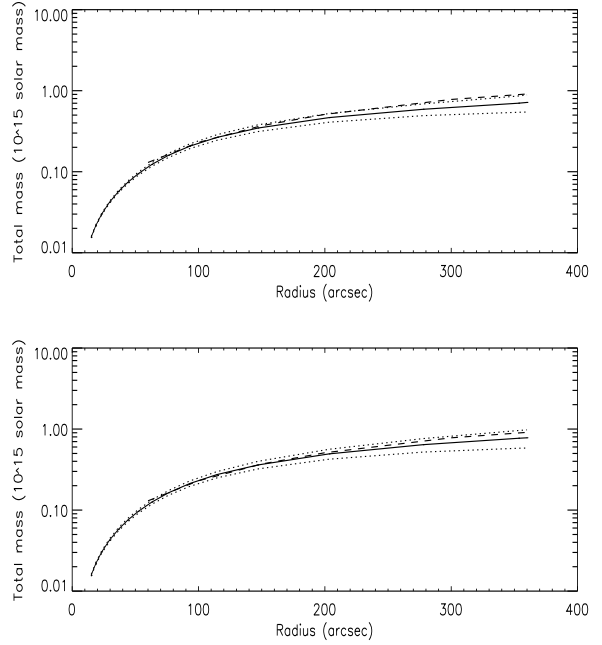,width=8cm,height=9cm}}
\caption{The projected mass profile in the universe model: $H_0$ = 100 
kms$^{-1}$Mpc$^{-1}$, $\Omega_m$ = 0.3, $\Omega_{\Lambda}$ = 0.7 (shown as the 
solid line; the dotted lines mark the confidence level 
of 68\%), and compared with that of Clowe \& Schneider (shown as the dashed 
line). In the upper panel, the solid line presents the projected mass within 
$6'$; the lower panel is for the projected mass within $15'$. }
\end{figure}

\subsection{Mass within the optical lensing arc}
The optical observation (taken on 27 February 1998 with the Canada France 
Hawaii Telescope, CFHT) shows an arc $\sim$ 153 kpc from the center of Abell 
1835 which is thought to be a gravitationally lensed image of a galaxy far 
behind Abell 1835 (see Fig.2 of Schmidt et al. 2001). We estimate the X-ray
projected mass inside this arc: $M_{arc}=0.80\pm0.02\times10^{14}$ M$_{\odot}$
(errors are at 68\% confidence level).
 
Since there is a two-temperature structure in the central part, we also
calculate the projected mass in the 2T model: assuming that the two temperature
components are in pressure equilibrium. Through the two-temperature model
fitting, we obtained the temperature and $norm$ for each component. Then using 
Eq. (6) the electron density $n_e$ of each can be derived. Because of 
the pressure equilibrium, we can calculate the projected mass from any 
component by using Eq. (9), and the result is about $0.7\times10^{14}$ 
M$_{\odot}$, which is still in agreement with that from the 1T model. 

Allen et al. (1996) calculated the lensing mass inside the radius of this 
arc, and the result is between 1.4$\times10^{14}$ M$_{\odot}$ and 
2$\times10^{14}$ M$_{\odot}$ for the redshift of the arc $0.6<z_{arc}<3.0$. 
Schmidt et al. (2001) also calculate the X-ray projected mass inside 
this arc using the {\it Chandra} data and their value is 
1.06$\sim$1.27$\times10^{14}$ M$_{\odot}$. Both results are larger 
than ours. We will discuss the reasons in the following section.

\subsection{The discrepancy between the X-ray mass and the optical 
lensing mass}
Sect. 4.4 shows that the projected mass within the optical lensing arc we
calculated is not consistent with previous results.
Our estimate is lower than that derived from the
analysis of Chandra data in Schmidt et al. (2001), the main reason being
probably the large difference 
of the temperatures determined by {\it XMM-Newton} and {\it Chandra}, which 
directly causes the difference of the mass. The X-ray gas temperature 
determined by {\it Chandra} is about 12 keV, but it is only about 7.6 keV for 
{\it XMM-Newton}.

On the other hand, Bartelmann (1995) shows that on average the cluster mass 
required for large arcs will be lower by a factor of $\sim$ 1.6 than expected 
from radially symmetric models, and that there is a probability of $\sim$ 20\% 
of overestimating the actual cluster mass by a factor of $\sim$ 2. The inner 
part of the X-ray isocontours of Abell 1835 is elliptical with an axis ratio 
of \begin{math}f=0.85\end{math} (see Fig.4 in Schmidt et al. 2001). Therefore, 
one reason for the difference between our result and the lensing mass might 
be the assumption of a symmetric model when calculating the lensing 
mass. Another reason may be the existence of radio structure (Ivison et al. 
2000) in the center region of Abell 1835 like in PKS 0745-191 (Chen et al. 2003), 
although the radio emission of the latter is much stronger. The radio 
plasma may fill a larger volume in the central region of the cluster or 
provide some additional pressure to support the X-ray gas. Therefore, when we 
consider the radio component, the projected mass may be larger and may be 
consistent with that of the optical lensing.

\section{Conclusion}
We have presented a detailed analysis of about 26 ksec of {\it XMM-Newton} 
observations of the galaxy cluster Abell 1835. Through the deprojected spectra 
analysis we derived the deprojected temperature profile, which is flat in the 
outer regions but decreases towards the center. The abundance is higher in the 
cluster center, which may be caused by the cD galaxy (Makishima et al. 2001). 
We also fit the spectrum in the central region with a two-temperature 
model, and the fit becomes better. It was found that the two temperature 
components coexist in the cluster center which is due to the gas cooling and/or 
the ISM associated with the cD galaxy (Makishima et al. 2001). The lower 
temperature component also has a lower abundance, which may be due to the fact 
that the metal-rich gas cools faster when temperature falls below about 2 keV 
(Fabian et al. 2001). Then considering a cooling flow model, we find that the 
intrinsic absorption of Abell 1835 is very small if not zero and the cooling 
flow rate is 656.2$^{+403.4}_{-360.2}$ M$_{\odot}$yr$^{-1}$, which is 
much smaller than that derived by the spatial method of $\sim$ 1600 M$_{\odot}$
yr$^{-1}$. This implies that there should exist some other energy sources that 
can heat the gas and prevent it from cooling down. Furthermore, we showed that 
the cooling flow model without a Mekal component cannot explain the data 
satisfactorily. Therefore, with any acceptable model which we investigated in 
Sect. 3, the main component in the cluster center (r $<0.75'$) is always a 
locally isothermal component and there should also exist another minor 
component, which may be a single phased cool component or a multiphased one. 
However, with the current data we cannot discriminate between these two 
models.

We fit the electron density profile by a double-$\beta$ model, and use the best 
fitting parameters to calculate the total mass. We find that within a radius 
of $6'$, $M_{tot}$ = 1.05$\pm$0.13$\times$ 10$^{15}$ M$_{\odot}$, which is 
consistent with the value derived by Majerowicz et al. (2002) within the error 
bars. We also calculate the total projected mass using a $H_0$ = 100 km 
s$^{-1}${Mpc}$^{-1}$, $\Omega_m$ = 0.3, $\Omega_{\Lambda}$ = 0.7 cosmological 
model and compare it with what was calculated in Clowe \& Schneider (2002) by 
the weak lensing method. We find that the X-ray projected mass within $6'$ is 
slightly lower than the weak lensing mass. However, if we assume that our 
best-fit $n_e(r)$ and $T(r)$ can be extrapolated out to 15' where mass 
measurements from Clowe \& Schneider (2002) are available, we show that the two 
determinations are consistent within the error bars.

Lastly, we calculate the projected mass within the optical lensing arc 
($r \sim$ 153 kpc) using the single temperature model, which is 
$0.80\pm0.02\times10^{14}$ M$_{\odot}$ and is almost the same as that from the 
double temperature method. This value is about half of the gravitational 
lensing mass. Our result is similar to that of another large cooling flow 
cluster PKS 0745-191 (Chen et al. 2003). Therefore, our results imply that the 
mass discrepancy between the X-ray and the optical lensing does exist in the well 
relaxed clusters. This discrepancy may be due to (i) the assumed symmetry
adopted in calculating the lensing mass, and (ii) a pressure support by the 
radio plasma in the cluster center.

\begin{acknowledgements}
We would like to thank the referee, Stefano Ettori, for the careful reading and useful remarks.
This research is partially supported by the Special Funds
for Major State Basic Research Projects and the National Natural Science
Foundation of China.
\end{acknowledgements}


\end{document}